\documentclass[aip,showpacs,reprint,groupedaddress]{revtex4-1}

\usepackage{graphicx}
\usepackage{dcolumn}
\usepackage{inputenc}
\usepackage{tabularx}
\usepackage{natbib}
\usepackage{setspace}
 \usepackage{fullpage}

\bibpunct{(}{)}{;}{a}{,}{,}

\begin{document}
\begin{center}
\huge{Quantitative High-Resolution Transmission Electron Microscopy of Single Atoms}
\end{center}

\begin{center}
B. Gamm$^{1*}$, H. Blank$^{1}$, R. Popescu$^{1}$, R. Schneider$^{1}$, A. Beyer$^{2}$, A. G\"olzh\"auser$^{2}$ and  D. Gerthsen$^{1}$
\end{center}

\noindent 
$^{1}$ Laboratorium f\"ur Elektronenmikroskopie, Karlsruher Institut  f\"ur Technologie (KIT), 76128 Karlsruhe, Germany

\noindent 
$^{2}$ Physik Supramolekularer Systeme, Universit\"at Bielefeld,  33501 Bielefeld, Germany

\noindent 
${*}$ Email: gamm@kit.edu

\noindent 
Keywords: single-atom imaging, quantitative HRTEM, phase contrast, scattering factors

\noindent
Running Title: Quantitative HRTEM of Single Atoms

\section*{ABSTRACT}
Single atoms can be considered as the most basic objects for electron microscopy to test the microscope performance and basic concepts for modeling image contrast. In this work high-resolution transmission electron microscopy was applied to image single platinum, molybdenum and titanium atoms in an aberration-corrected transmission electron microscope. The atoms are deposited on a self-assembled monolayer substrate which induces only negligible contrast. Single-atom contrast simulations were performed on the basis of Weickenmeier-Kohl and Doyle-Turner form factors. Experimental and simulated image intensities are in quantitative agreement on an absolute intensity scale which is provided by the vacuum image intensity. This demonstrates that direct testing of basic properties like form factors becomes feasible. 


\section*{INTRODUCTION}
Imaging single atoms has always been a challenging goal in electron microscopy which is interesting with respect to studying single-atom dynamics in context with cluster nucleation and coarsening in application fields like catalysis and crystal growth. Apart from application relevance, single atoms can be considered as the most basic objects in electron microscopy. Imaging single atoms by electron microscopy has been previously demonstrated by scanning transmission electron microscopy (STEM) for some time already \citep{crewe1970,batson2002,Krivanek2010}. Single atoms are visualized in the annular dark-field STEM mode due to the small electron probe formed by the condenser-lens system and the atomic-number dependent scattering-angle distribution. 

Alternatively, high-resolution transmission electron microscopy (HRTEM) can be applied for single-atom imaging \citep{iijima1977,Ohnishi1998,meyer2008}. HRTEM is based on phase contrast where the interaction between electron wave and atom only induces a phase shift with negligible amplitude variation in the image electron wave. This leads to weak single-atom contrast which depends strongly on the defocus and other lens aberrations. Comparison between experimental and simulated images taken at different defocus values is therefore indispensable for the interpretation of HRTEM images. This was shown by several groups \citep{iijima1977,Koizumi2001} who compare characteristic features of experimental and simulated single-atom images or relative intensities between atoms in an image. 

However, the comparison must be performed on an absolute intensity scale to ultimately analyze microscope performance, basic concepts for modeling image formation or even form factors. Absolute intensity scaling can only be achieved by image normalization with the vacuum intensity rather than intensity normalization with respect to another sample feature. In this sense, we present for the first time 
quantitative agreement between a series of experimental and calculated single-atom images on an absolute intensity scale. The agreement is found for single exposure images, and no averaging over several images is needed to enhance the single-atom contrast. Our work is facilitated by an correction-lens system for spherical and other aberrations (for short C$_{S}$-corrector) which improves the resolution to better than 0.1 nm and allows to measure precisely residual lens aberrations. 

\begin{table}
 \caption{\label{tab:tab1} Measured residual aberrations for the Pt-specimen images (Z: defocus, C3, C$_{S}$: spherical aberration, A1: twofold astigmatism, A2: threefold astigmatism, A3: fourfold astigmatism, A4: fivefold astigmatism, B2: axial coma, S3: star aberration).}
\begin{tabularx}{0.45\textwidth}{c|c|c|c}
\hline\hline
Z&C3,C$_{S}$&B2&S3\\
-19 to +16 nm&804.5 nm &8.435 nm&84.93 nm \\
 & &$24.7^{\circ}$&$170.5^{\circ}$\\
\hline
A1&A2&A3&A4\\
273.2 pm &12.07 nm &369.7 nm &15.29 $\mu$m \\
$-80.3^{\circ}$&$-119.3^{\circ}$&$175.1^{\circ}$&$-60.4^{\circ}$\\
\hline\hline
\end{tabularx}

\end{table}  

\section*{MATERIALS AND METHODS}
The samples were prepared by electron-beam evaporation of approximately 0.2 monolayers Pt (1 monolayer: $1.3\cdot 10^{15} atoms/cm^{2}$), Mo or Ti, on a self-assembled monolayer (SAM) film denoted as nanosheet in the following. It consists of 1,1'-biphenyl-4-thiol molecules (C$_{12}$H$_{10}$S) with a thickness of about $1.0~nm$ which are crosslinked by electron irradiation \citep{turchanin2009}. HRTEM was carried out with an aberration-corrected FEI Titan$^{3}$ 80-300 microscope operated at 300 keV. A defocus series was acquired which consists of 36 single images with a defocus step of $1~nm$ between $19~nm$ overfocus and $-16~nm$ underfocus. The exposure time was $1~s$. The C$_{S}$-corrector was used to minimize and determine the residual objective-lens aberrations on the basis of the procedure described by Uhlemann and Haider \citep{uhlemann1998}. Tab.\ref{tab:tab1} shows the aberration values which were used for the image simulations in the case of the Pt-atom images. 

\begin{figure}
\includegraphics[width=0.45\textwidth]{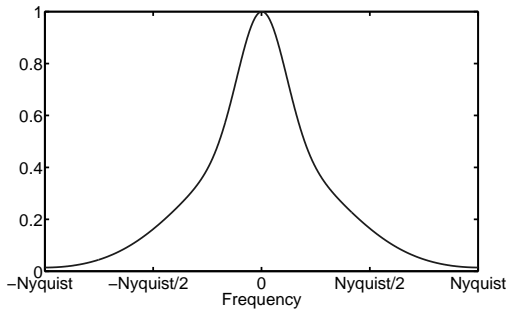} 
\caption{\label{fig:fig2mtf} Modulation transfer function of the Gatan Ultrascan 1000 CCD-camera determined by the edge method. }
\end{figure}

\begin{figure*}[ht]
\includegraphics[width=0.9\textwidth]{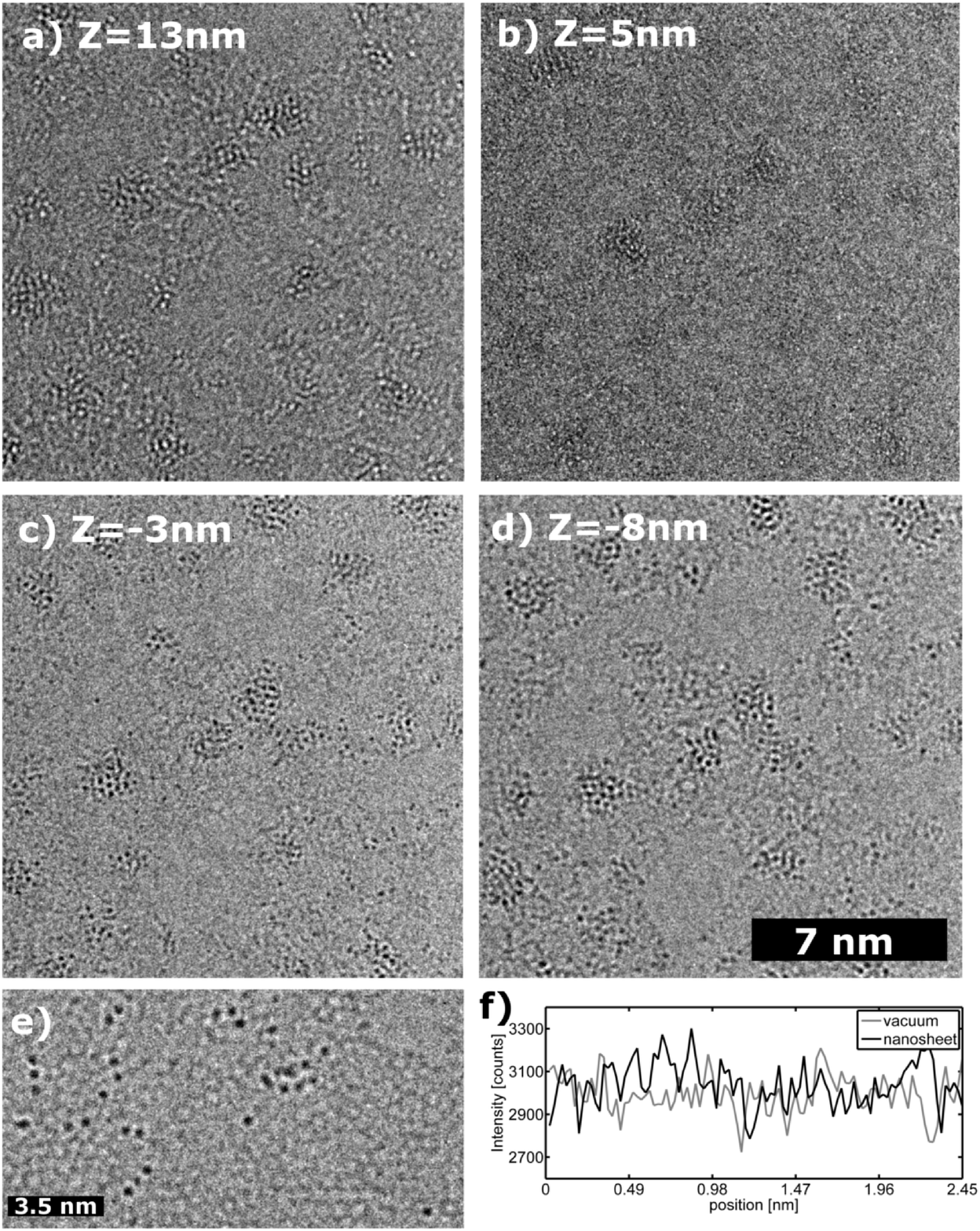}
\caption{ \label{fig:fig1} Experimental HRTEM images and substrate contrast. a)-d): HRTEM images of single Pt-atoms and small clusters of Pt-atoms on a nanosheet acquired with a FEI Titan$^{3}$ 80-300 microscope at 300 kV and different defocus values $Z$ as indicated. e) Magnified and strongly defocused ($Z = -19~nm$) image. Single Pt-atoms and substrate are clearly visible. f) Intensity profiles taken from a HRTEM image d) taken at $Z=-8~nm$ in a vacuum (gray) and nanosheet (black) region. Intensity (absolute counts) and noise levels are comparable.}
\end{figure*}

Image simulations are based on an object exit wave function $\psi_{O}$ calculated with the multislice formalism \citep{cowley1957} of the STEMsim program \citep{stemsim}. A region around a single Pt-atom was defined large enough to avoid artifacts that arise from the periodic boundary conditions implied in the multislice algorithm. The atomic potential is calculated by inverse Fourier transformation from Weickenmeier-Kohl \citep{weickenmeier1991} or Doyle-Turner \citep{doyle1968} form factors. The former explicitly deals with absorption arising from  inelastic scattering. The aberrations of the imaging-lens system of the microscope modify the phase of $\psi_{O}$ as a function of the spatial frequency $g$ which is described by the wave transfer function $T(g)$. The image wave function $\psi_{i}$ is given in Fourier representation by $\psi_{i}=\psi_{O} \cdot T(g) = \psi_{O} \cdot exp(-i \chi (g))$with the spatial frequency-dependent phase shift $\chi (g)$ elaborated up to the fifth order by Uhlemann \citep{uhlemann1998} and Kirkland \citep{kirkland2006}. The partial spatial coherence of the incident electron wave is taken into account by multiplying $T(g)$ with a damping envelope function \citep{wade1977}, which is negligible for the small C$_{S}$ and $Z$ values used in this work. Partial temporal coherence is characterized by the parameter defocus spread $\Delta$ which is $\approx 4.5~nm$ for our electron microscope. The effect of partial temporal coherence was taken into account by simulating images for nineteen different defocus values around the chosen $Z$ value and weighting the different contributions to the image according to a Gaussian distribution with a FWHM given by $\Delta$. The FEI Titan$^{3}$ 80-300 is equipped with a monochromator which, however, was not excited in the present work. Slight illumination inhomogeneities may occur due to an additional solid aperture implemented in the gun section of the microscope. Image detection by the CCD camera dampens pixel-size contrast like a low-pass filter. This behavior is described by the modulation transfer function (MTF) which was measured by the edge method described by \cite{weickenmeier1995} and taken into account by convoluting the image with the MTF. The MTF for the Gatan Ultrascan 1000 CCD camera is shown in Fig.\ref{fig:fig2mtf}. It was determined from images taken under the same experimental conditions than the images presented in this work. Nyquist frequency corresponds in the case of Pt-atom images to $\approx 20~nm^{-1}$.

\section*{RESULTS}

\begin{figure}
\includegraphics[width=0.45\textwidth]{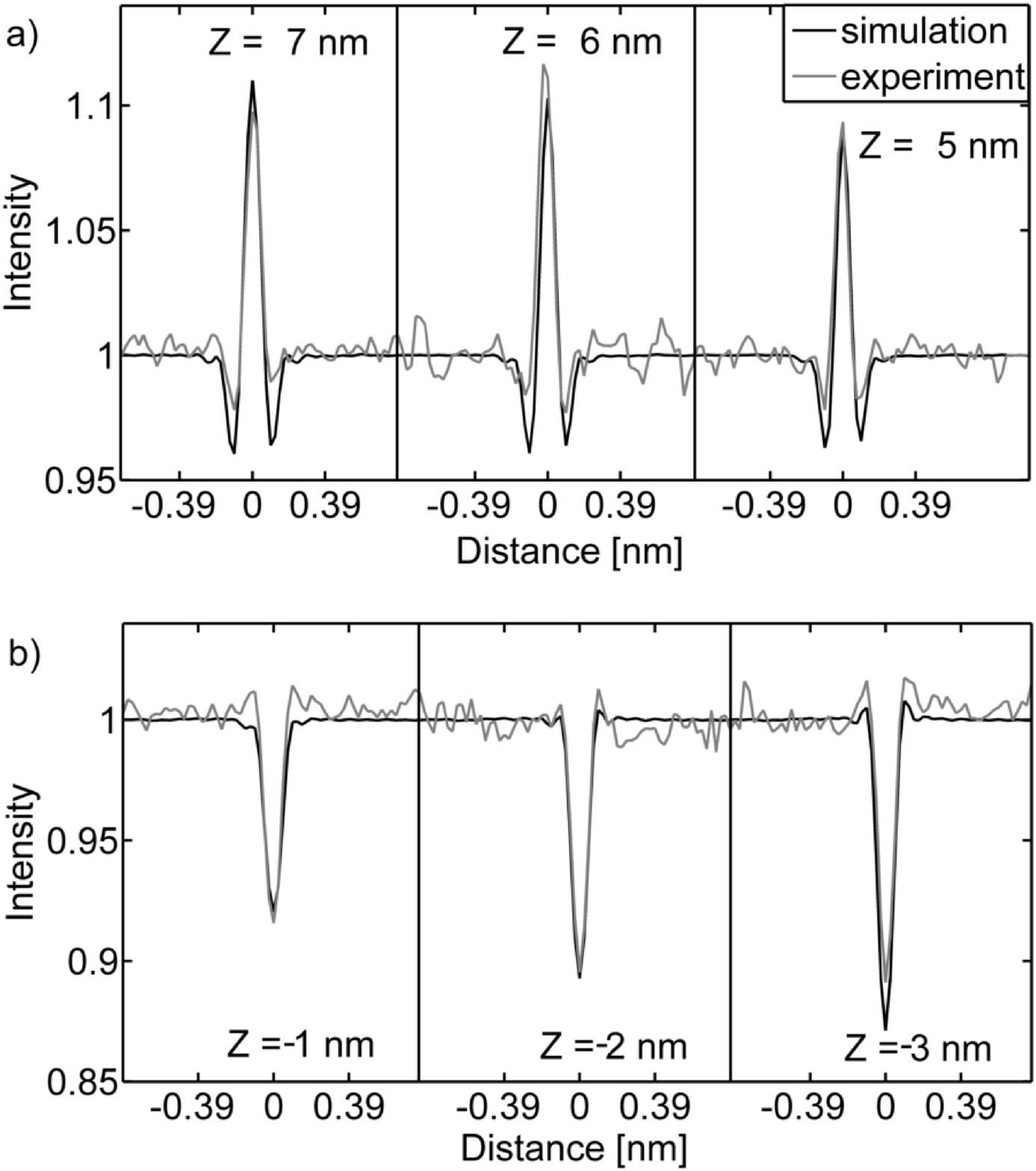} 
\caption{\label{fig:fig2} Comparison of simulated and experimental intensity profiles. Experimental (gray) and simulated (black) intensity profiles across single Pt-atoms at a) overfocus values of 7 nm, 6 nm and 5 nm and b) underfocus values of -1 nm, -2 nm and -3 nm. }
\end{figure}

Fig. \ref{fig:fig1} shows typical HRTEM images of small planar Pt-clusters and isolated Pt-atoms on a nanosheet at different  $Z$-values as indicated on the images. The spatial sampling rate was $0.0245~nm/pixel$. Fig. \ref{fig:fig1}a) shows bright atom contrast under overfocus conditions. The intensity of the Pt-atom contrast vanishes almost completely at $Z=5~nm$ (Fig. \ref{fig:fig1}b)), which is the image with the lowest contrast that could be quantifiied. At a defocus value of $Z=2-3~nm$ no contrast is observed. The minimum is shifted into the overfocus region due to the finite positive value of $C_{S}$.  Contrast is reversed under underfocus conditions (Fig. \ref{fig:fig1}c,d)) where dark atom contrast is observed. 
The atoms are highly mobile on the nanosheet which can be inferred from the change in the atom configuration of the clusters. Fig. \ref{fig:fig1}e) shows a magnified region where isolated atoms can be discriminated from atoms in small clusters. Furthermore a comparably large defocus of $Z=-19~nm$ allows distinguishing the contrast associated with the nanosheet from the vacuum as shown in Fig. \ref{fig:fig1}e). The top right corner shows a small hole in the nanosheet marked by an arrow. Fig. \ref{fig:fig1}f) presents intensity line profiles measured at $Z=-8~nm$ along a line across a vacuum and nanosheet region with mean intensities corresponding to 2994 and 3038 counts. Almost the same standard deviation of the intensity (90 counts for vacuum and 110 counts for the substrate) for these two regions indicates the uniform small thickness of the nanosheet and its negligible influence on image contrast at small defocus values. The analysis of the intensity profiles yields a noise level of about 3 \%. Therefore, the nanosheet contrast can readily be neglected in the simulations for the defocus range observed. Normalization of the experimental images with the nanosheet background intensity corresponds in a good approximation to normalization with the vacuum intensity. Nonetheless, a region in direct vicinity of the analyzed atom should be taken for normalization. Due to the monochromator of the microscope, the illumination is not uniform over the field of view which can be seen from the gradient in the images in Fig. \ref{fig:fig1}a)-d). For several different defocus values intensity profiles of isolated atoms were obtained from experimental images by averaging profiles along the horizontal and vertical image axis. Fig. \ref{fig:fig2} shows representative experimental intensity profiles at different defocus values. Intensity profiles of simulated single-atom images were extracted in the same way and are also contained in Fig. \ref{fig:fig2}. Experimental and simulated profiles agree well with respect to the noise level of 3 \%. For quantitative comparison of experimental and simulated images, the maximum intensity normalized with respect to the background intensity and the full width at half maximum (FWHM) of the peak were chosen. Another quantitative feature is the radius of the first side minimum/maximum (overfocus/underfocus images) of the fringe-like contrast around the central peak.

Simulated data for the peak intensity values, FWHM and distance (radius) from the center to the first side minimum/maximum (underfocus/overfocus image) are listed in Table \ref{tab2} together with the experimental data for twelve different $Z$ values. Within the signal-to-noise ratio full agreement between simulated and experimental peak intensities is obtained for Weickenmeier-Kohl form factors. For Doyle-Turner form factors, a different defocus dependence is observed as shown in the forth column of Table \ref{tab2}.

\begin{table*}[ht]
 \caption{\label{tab2} Comparison of contrast characteristics between experimental and simulated images of single Pt-atoms at 12 different defocus values (1st column). Columns 2-4: peak contrast of single Pt-atoms with respect to the background intensity in percent for experimental und simulated data for Weickenmeier-Kohl (WK) and Doyle-Turner (DT)form factors. Column 5-8: FWHM of central peak and radius of the first side minimum (overfocus images) and first side maximum (underfocus images) ring in experiment and simulation. The error for distance measurements is approximately 0.013 nm corresponding to half a pixel.}
\begin{tabularx}{\textwidth}{c|ccc|cc|cc}
\hline\hline
Defocus&Peak&Peak&Peak&FWHM&FWHM&Min&Min\\
&&Sim WK&Sim DT&[nm]&Sim [nm]&[nm]&Sim [nm]\\ \hline
18 nm&11.8\%&9.8\%&9.1\%&0.135&0.147&0.172&0.147\\
14 mn&13.7\%&10.3\%&9.7\%&0.135	&0.135&0.147&0.147\\
10 nm&12.0\%&11.1\%&9.9\%&0.123	&0.123&0.123&0.110\\
7 nm&9.8\%&11\%&9.1\%&0.098&0.098&0.098&0.098\\
6 nm&11.7\%&10.3\%&8.5\%&0.098&0.098&0.098&0.098\\
5 nm	&9.3\%&9 \%&7.7\%&0.086&0.074&0.074&0.074\\
-1 nm&-9.2\%&-8.6\%&-1.8\%&0.086&0.086&0.098&0.098\\
-2 nm&-11.7\%&-12.1 \%&-2.8\%&0.098&0.086&0.098&0.098\\
-3 nm&-12.2\%&-14.9\%&-4.7\%&0.098&0.098&0.098&0.110\\
-6 nm&-13\%&-18.4\%&-9.5\%&0.110&0.110&0.123&0.123\\
-10 nm	&-15\%&-16.4\%&-12.4\%&0.123&0.123&0.123&0.147\\
-15 nm	&-13.2\%&-13.7\%&-12.3\%&0.147&0.147&0.172&0.172\\
\hline\hline
\end{tabularx}
\end{table*}

For Weickenmeier-Kohl form factors good agreement is also found for the radius of the first side minimum/maximum as a result of the precise determination of objective lens aberrations. Quantitative agreement between simulation and experiment is also demonstrated in Fig. \ref{fig:fig2}. It shows six representative intensity profiles for three different overfocus values in Fig. \ref{fig:fig2}a) and three different underfocus values in Fig. \ref{fig:fig2}b). Peak intensity and FWHM are in good agreement. Overfocus profiles clearly show the fringe-like contrast around the central maximum. We emphasize that quantitative agreement between experimental and simulated single-atom contrast is achieved on an absolute scale only if the MTF of the CCD camera is taken into account. The importance of the MTF was also outlined in context with the "Stobbs-factor" problem \citep{thust2009}, where quantitative agreement for crystalline specimens is found when accouting for the MTF. 

\begin{figure}
\includegraphics[width=0.45\textwidth]{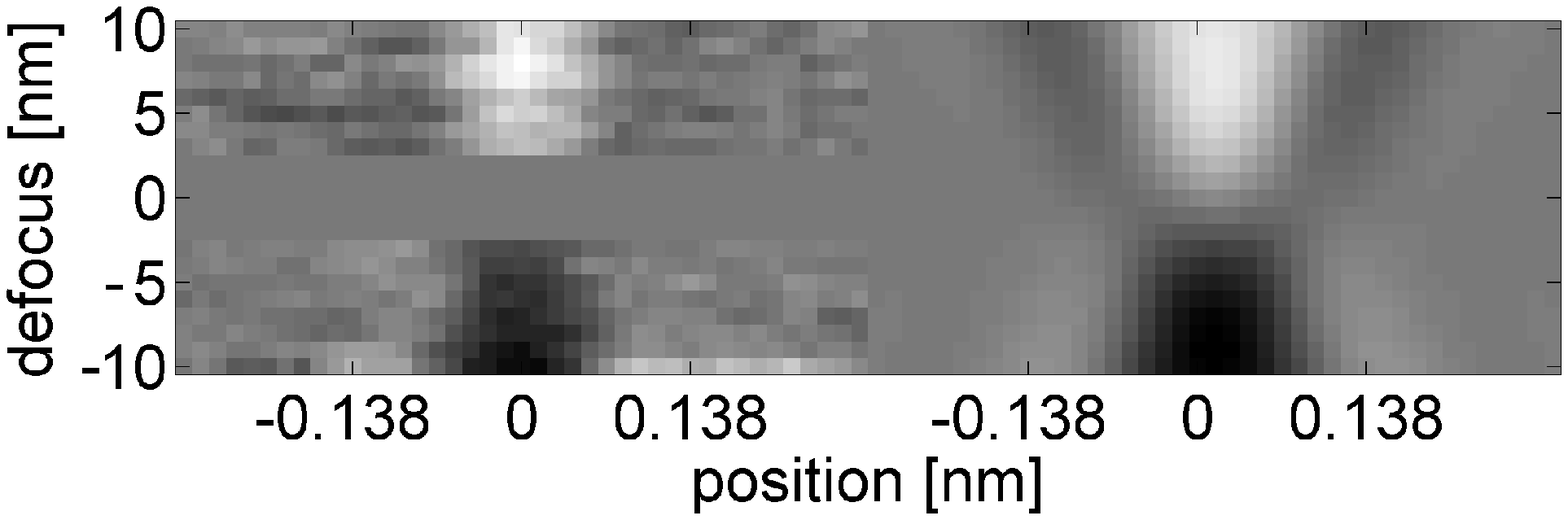}
\caption{\label{fig:fig3} Comparison of experimental and simulated intensity profiles of single Mo-atoms as a function of defocus. Gray-scale values of the images correspond to absolute image intensities. The left side shows horizontally intensity profiles stacked vertically for different defocus values between $Z=10~nm$ and $Z=- 10~nm$. The missing part at defocus values around $Z=0~nm$ is due to the low contrast of Mo-atoms. The right side of the image shows the corresponding simulated intensity profiles. Good agreement is achieved except for $Z=-6~nm$.}
\end{figure}

The method for quantitative single-atom imaging shown in this report is generally applicable to all elements which can be deposited in very small amounts onto a nanosheet by electron-beam evaporation. Apart from platinum (atomic number: 78) we also analyzed molybdenum (atomic number: 42) and titanium (atomic number: 22). Mo is a significantly weaker scatterer than Pt resulting in a reduced atom contrast which is more subject to noise. Nonetheless a quantitative comparison between experimental and simulated single-atom intensity profiles is possible. A defocus series of the Mo-sample was taken in the same way as for Pt apart from magnification that was increased to achieve a sampling rate of $0.0138~nm/pixel$. Due to the strong effect of the MTF, the contrast of Mo-atoms cannot be directly compared to the contrast of Pt-atoms, and should only be compared with image simulations performed under the same conditions (i.e. aberrations, magnification). From the experimental defocus series intensity profiles for isolated single Mo-atoms were taken in analogy to Fig.\ref{fig:fig2} for a defocus range from $Z=-10~nm$ to $Z=+10~nm$. The profiles were then stacked in y-direction into an image which is shown in Figure \ref{fig:fig3}. The left part of the image shows the experimental profiles. The right side shows profiles from simulated defocus-series images. The missing defocus values in the experimental data are due to the low contrast of Mo-atoms in these  images. Apart from defocus value $Z=6~nm$ atom contrast is in good qualitative and even quantitative agreement. In the underfocus regime, the bright halo around the dark atom contrast is reproduced as observed in the simulation. This feature corresponds to the first side maximum already discussed earlier.

\begin{figure}
\includegraphics[width=0.45\textwidth]{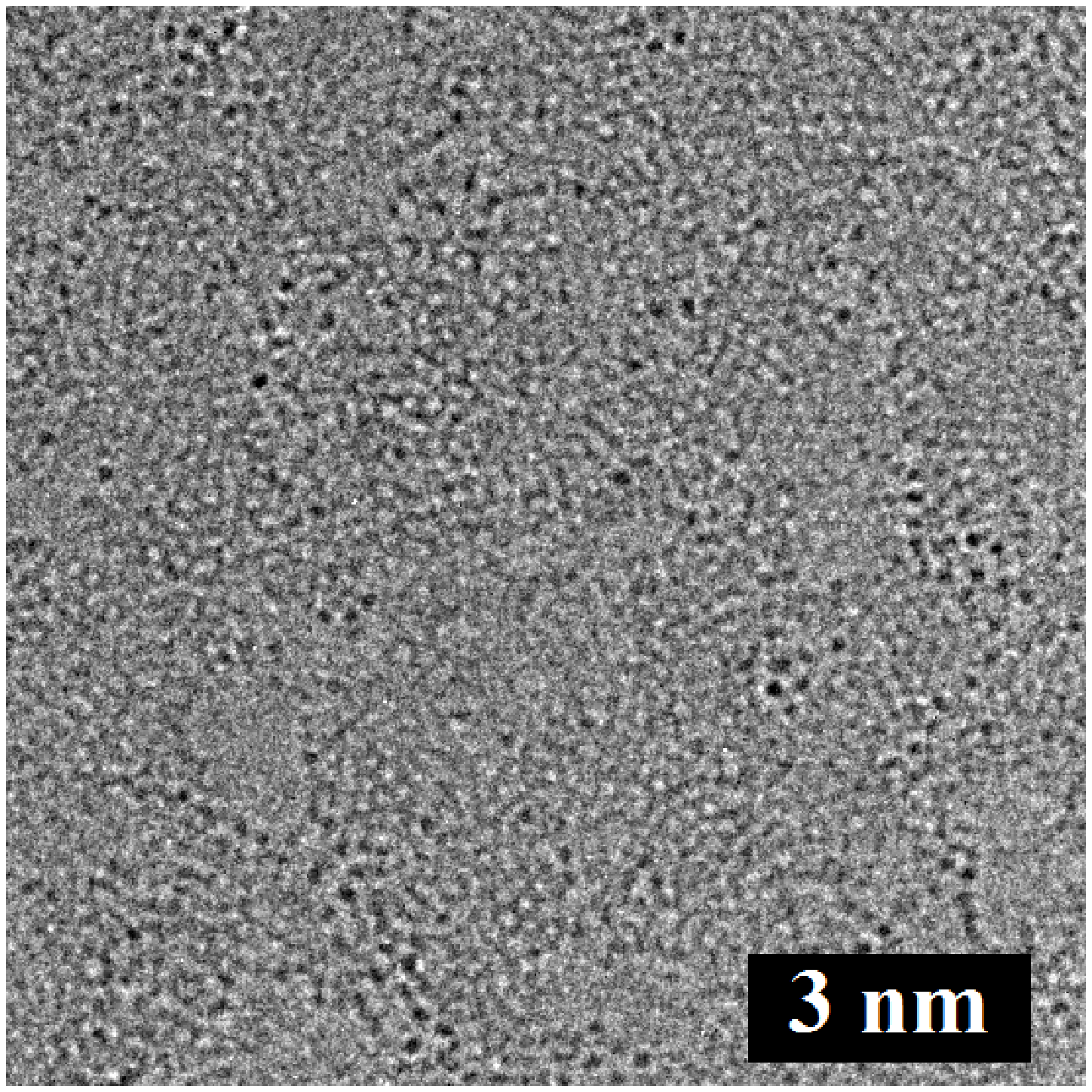}
\caption{\label{fig:fig4} Experimental HRTEM image of deposited Ti-atoms on a nanosheet, imaged with a defocus value of $Z=-10~nm$. Isolated Ti-atoms are clearly visible due to dark atom contrast. The contrast of nanosheet is strong with respect to vacuum intensity and can therefore not be neglected.}
\end{figure}

In the case of Ti the comparison becomes more difficult mainly due to the nanosheet contrast, which hampers the possibility to find isolated atoms and to normalize the contrast with respect to vacuum intensity. Figure \ref{fig:fig4} shows Ti-atoms on the nanosheet at a rather high defocus of $Z=-10~nm$. Due to the comparably weak scattering power of the Ti-atoms, the Ti-contrast does not significantly exceed the contrast of the nanosheet anymore. This prevents normalization of the atom contrast with respect to the background intensity.
For smaller defocus values single-atom contrast is even further  diminished but the noise remains the same. It should also be noted that the binding to other atoms can slightly change image contrast for lighter elements \citep{meyer2011}. Based on the quantification of Mo-atom contrast and the comparison to Ti-atoms we conclude that the lightest elements, which allow quantification of single-atom contrast, should be elements in the range of atomic number 30. This is mainly due to the increasing contrast of the nanosheet with respect to the single-atom contrast, but also due to the reduction of the signal-to-noise ratio which can be achieved in a single-exposure image of light single atoms. 
Furthermore we note, that each biphenyl molecule contains a S-atom (atomic number: 16) which cannot be distinguished from Ti-atoms (atomic number: 22). However the nanosheet may contain only a small concentration of S-atoms due to the nanosheet preparation procedure \citep{turchanin2009}.

\section*{DISCUSSION}
In the previous section, quantitative agreement between experimental and simulated single-atom contrast was shown. Negligible contrast of the substrate is essential for normalization of single-atom contrast with respect to vacuum intensity. This is a prerequisite to quantitatively compare experimental and simulated single-atom intensities on an absolute scale and to test form factors which are used in all the image simulation software packages. Weickenmeier-Kohl (WK) and Doyle-Turner (DT) form factors are established for high-energy electron diffraction and are therefore tested in this work for  wave-function simulations. Elastic scattering is described by a real form factor $f_{el}$ while an imaginary (absorptive) part $f_{abs}$ was introduced by Weickenmeier and Kohl (1991) to take into account thermal diffuse scattering in crystals. A better agreement between simulated and experimental images of single Pt-atoms is obtained for the WK form factor which is surprising on first sight because thermal diffuse scattering should be irrelevant for single atoms. This peculiarity can be understood by considering in detail the calculation of $f_ {el}$ for the WK and DT form factors. Although the physics behind $f_{el}$ is identical, the fit functions for $f_{el}$ differ for the WK and DT form factors. The elastic part of the WK form factors is calculated on the basis of  $f_{el}(s)=s^{-2}\sum_{i=1}^6 A_{i} [1-exp(-B_{i}s^{2})]$ as a function of $s=g/ 4 \pi$ with parameters $A_{i}$ and $B_{i}$ given by Weickenmeier and Kohl (1991). Doyle and Turner (1968) suggested $f_{el}(s)=\sum_{i=1}^5 A_{i} exp(-B_{i}s^{2})$ with different parameters $A_{i}$ and $B_{i}$ which are listed, e.g., in the International Tables for Crystallography \citep{cryst1992}. The different fit functions and parameters $A_{i}$, $B_{i}$ lead to higher scattering amplitudes at small spatial frequencies for $f_{el,WK}$ which is the main reason for the good agreement of the WK form factors with the experimental data. The influence of the imaginary part of the WK form factor is negligible for single-atom scattering. In both cases, the form factors are dampened by a Debye-Waller exponent $exp(-M g^{2})$ which contains the Debye Waller factor M. However, the variation of M between, e.g., $0~\mathring{A}^{2}$ and $0.046~\mathring{A}^{2}$ \citep{sears1991} only leads to a negligible reduction of the amplitude of the scattered electron wave for a single Pt-atom. Since the argument of the exponential function is proportional to $-g^{2}$ the influence of M is negligible for the small g-values which are important for single-atom imaging.

While form factors are the basis for calculating electron-wave functions of an atom, image contrast depends also on instrumental parameters. These are magnification, objective lens aberrations, partial incoherencies as well as the modulation transfer function of the imaging detector. All these parameters must be known as precisely as possible during the experiment for quantitative studies. These are then used in the subsequent image simulation procedure. The $C_S$-corrector reduces spherical aberrations to a point where the main influence on image contrast is defocus. Variation of defocus allows analysis of the defocus dependence of image contrast. If image aquisition is performed with a CCD-camera, the MTF has a strong effect on image contrast, and quantitative agreement is only obtained if the MTF is determined for the camera and camera mode which is used for the experiment. Finally, partial coherence also has an effect on image contrast. Partial spatial coherence can be neglected, because primary objective lens aberrations are small due to the $C_S$-corrector. Partial temporal coherence was calculated from instabilities of high tension and lens current, which turned out to be precise enough for an agreement of image contrast.

\section*{SUMMARY}
To summarize, through-focal series of HRTEM images of single Pt-, Mo- and Ti-atoms were acquired with an aberration-corrected transmission electron microscope. The samples were prepared by electron-beam evaporation on self-assembled monolayer substrates wich provide negligible background contrast as a prerequisite for quantitative analysis of single-atom contrast on an absolute intensity scale. The imaging process is modeled by Fourier optics taking into account all known instrumental effects including wave aberrations, partial coherence and the modulation transfer function of the CCD camera. Doyle-Turner and Weickenmeier-Kohl form factors are compared for wave-function simulations. We find quantitative agreement between simulated and experimental single-exposure image intensity of single Pt- and Mo- atoms based on Weickenmeier-Kohl form factors which provide a more adequate modeling of the real part of the form factor compared to the Doyle-Turner form factors. The influence of Debye-Waller dampening and the absorptive part of the Weickenmeier-Kohl form factors is negligible for single-atom images. We emphasize that the agreement is found on an absolute intensity scale based on raw image data without manipulation of the images like filtering scaling or brackground substraction. Single-atom contrast of Ti-atoms with an atomic number of 22 is not sufficiently strong with respect to the nanosheet contrast. We estimate that the presented method allows single-atom contrast quantification down to an atomic number of approximately 30, the limiting factor mainly being signal-to-noise ratio.

\section*{ACKNOWLEDGMENTS}
This work has been performed within the project C4.5 of the DFG Research Center for Functional Nanostructures (CFN). It has been further supported by a grant from the Ministry of Science, Research and the Arts of Baden-W\"urttemberg (Az: 7713.14-300).


\bibliography{singleatoms.bib}

\end{document}